\documentclass[twocolumn,showpacs,floats,floatfix,superscriptaddress,aps,pra]{revtex4}
\usepackage{amsfonts}
\usepackage{amssymb}
\usepackage{amsmath}
\usepackage{graphicx}
\usepackage{bm}
\usepackage{color}
\usepackage{epsfig}
\usepackage{calc}
\usepackage{ifthen}

\begin{document}

\title{Electronic controllable broadband and robust terahertz surface plasmon-polaritons switch based on hybrid ITO waveguide coupler}
\date{\today }

\begin{abstract}
The surface plasmon-polaritons (SPPs) switch is the key element of the integrated devices in optical computation and terahertz (THz) communications. In this paper, we propose a novel design of THz SPPs switch based on quantum engineering. Due to the robustness of coherent quantum control technique, our switch is very robust against with perturbations of geometrical parameters and presents a good performance at on-state (and off-state) from 0.5 THz to 0.7 THz. The on-state and off-state of our device can be controlled by the external voltage. We believe this finding will be the great improvement for the integrated optical computing and THz communications. 
\end{abstract}

\author{Wei Huang}
\affiliation{Guangxi Key Laboratory of Optoelectronic Information Processing, Guilin University of Electronic Technology, Guilin 541004, China}

\author{Erxiang Dong}
\affiliation{Guangxi Key Laboratory of Optoelectronic Information Processing, Guilin University of Electronic Technology, Guilin 541004, China}

\author{Yu Cheng}
\affiliation{Guangxi Key Laboratory of Optoelectronic Information Processing, Guilin University of Electronic Technology, Guilin 541004, China}

\author{Songyi Liu}
\affiliation{Guangxi Key Laboratory of Optoelectronic Information Processing, Guilin University of Electronic Technology, Guilin 541004, China}

\author{Shijun Liang}
\affiliation{National Laboratory of Solid State Microstructures, School of Physics, Collaborative Innovation Center of Advanced Microstructures, Nanjing University, Nanjing, 210093 China}

\author{Yuping Yang}
\affiliation{School of Science, Minzu University of China, Beijing 100081, China}

\author{Shan Yin}
\email{syin@guet.edu.cn}
\affiliation{Guangxi Key Laboratory of Optoelectronic Information Processing, Guilin University of Electronic Technology, Guilin 541004, China}

\author{Wentao Zhang}
\email{zhangwentao@guet.edu.cn}
\affiliation{Guangxi Key Laboratory of Optoelectronic Information Processing, Guilin University of Electronic Technology, Guilin 541004, China}
\maketitle

%42.82.Et Waveguides, couplers, and arrays
%42.81.Qb Fiber waveguides, couplers, and arrays
%42.79.Gn Optical waveguides and couplers
%32.80.Xx Level crossing and optical pumping

%***************************************************************

\section{Introduction}

%***************************************************************
Terahertz (THz) Surface plasmon-polaritons (SPPs) is the most active research area in optical integrated device, due to the unique features of SPPs in relative smaller wavelength, energy confinement and field enhancement \cite{Liu2014, Williams2008}. 
THz SPPs devices are widely used in THz communications and THz information processing. Optical switch is the deice which can select the path of light propagation according to 'on-state' and 'off-state', which is the fundamental element in optical communications \cite{Mellette2016, Guo2019} and optical computing \cite{Seok2019, Yu2013,Hwang2017}. 
Combining with advantages of THz SPPs device and optical switch device, controllable THz SPPs switch has a large potential applications on THz communications and computation. 
Currently, there are few papers working on this idea, such as light-controlled THz SPPs device \cite{Cao2018} and THz switches based on graphene \cite{Luo2017}. 

Most recently, there are two remarkable papers which proposed broadband and robust THz SPPs waveguide coupler based on coherent quantum control \cite{Huang20201, Huang20191}. 
The technique of coherent quantum control can be broadly used in many-body physics \cite{Huang20192}, waveguide coupler \cite{Huang20193} and graphene SPPs waveguide coupler \cite{Huang2018}. 
Due to the robustness of coherent quantum control, the especial curved SPPs waveguide coupler \cite{Huang20201, Huang20191} can completely transfer the energy of SPPs from input to output waveguide with wide working band and insensitivity to the perturbations of geometrical parameters. 
However, the previous researches \cite{Huang20201, Huang20191} can not control the output path of SPPs.   
The light-controlled THz SPPs device \cite{Cao2018} is not robust against geometrical parameters and only can operate in the narrowband. The terahertz switch based on graphene \cite{Luo2017} can operate at broadband, however, their device is not robust against geometrical parameters and this design of 'on-state' (or 'off-state') propagates (or not propagates) along with single SPPs waveguide. Thus, their device can not work as optical switch array for optical computing and can not controllable switch optical path. 
Besides, the fabrication of graphene sheet is much complex comparing to our design. 

\begin{figure*} [htbp]
\centering
\includegraphics[width=0.7\textwidth]{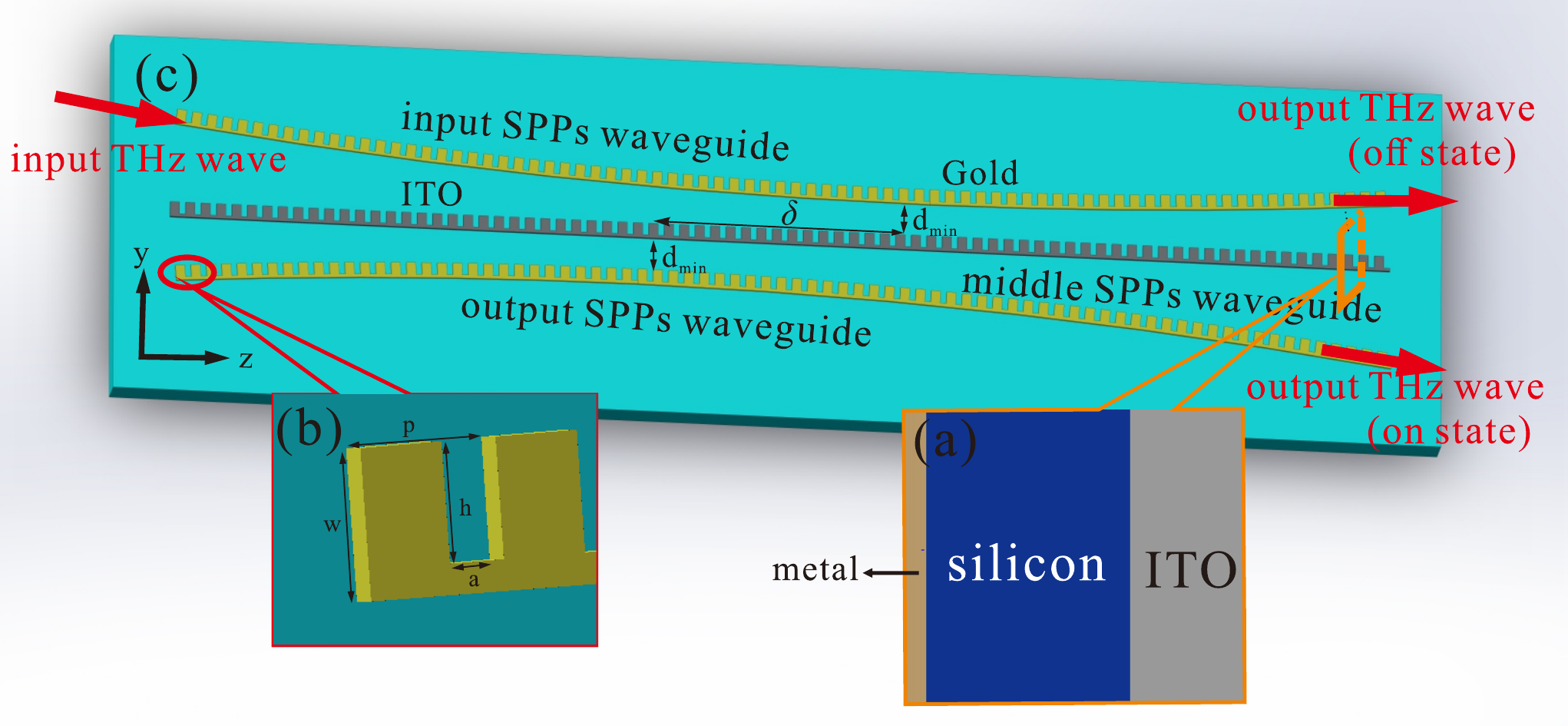}
\caption{The schematic figure of our design. (a) The cross-section view of middle SPPs waveguide with sandwich structure. (b) The detailed structured view of input/middle/output SPPs waveguide. (c) The overview of special curved configuration of three SPPs waveguide coupler. } 
\label{fig1}
\end{figure*}

In this paper, we propose the device based on especial curved SPPs waveguide coupler inserted the hybrid indium tin oxide (ITO) waveguide in the middle of two metallic waveguides, as shown in Fig. ~\ref{fig1}, where the top waveguide functions as the input port and output port in 'off-state' and the bottom one is the output port in 'on-state'. 
The hybrid ITO waveguide is the sandwiched structure with $180$ $nm$ ITO layer, $SiO_{2}$ layer and metallic layer with several hundred nanometers.
When we apply the external voltage on the metallic layer, ITO layer with very low conductivity can be considered as the insulating layer \cite{Li2019}. 
When the external voltage is not apply on the metallic layer, the ITO layer has high conductivity and can hold SPPs in THz region as the same as sliver and gold \cite{Franzen2008}. 

Due to this unique feature of the ITO layer, we can apply the external voltage to control the conductivity of ITO corresponding to 'on-state' and 'off-state'.  
The 'on-state' without applying external voltage (ITO layer with high conductivity) can be considered as the three SPPs waveguide coupler. 
Thus, the energy of SPPs can completely transfer form input to output SPPs waveguide.
Our device on the 'off-state' with applying external voltage (ITO layer with very low conductivity) can be consider as two SPPs waveguide coupler without middle SPPs waveguide. Therefore, the distance between input and output waveguide becomes much larger and coupling strength between input and output waveguide turns much lower. The energy of SPPs confines in the input waveguide along with propagation.

In this paper, we firstly propose a typical example design (see Fig.~\ref{fig1}) for our SPPs switch coupler and we illustrate the function of our switch according to 'on-state' and 'off-state' based on numerical calculations of SPPs coupler equation, as shown in Fig.~\ref{fig2}. In section III, we further proof that our device is the robust against perturbations of geometrical parameters ($L$, $R$, $\delta$ and $d_{\text{min}}$) for both 'on-state' and 'off-state' (see Fig.~\ref{fig3}). Besides, we demonstrate that our switch can operate at broad bandwidth, as shown in Fig.~\ref{fig4}.

\section{Model and Device}

Our device consists with three SPPs waveguide with special curved input and output SPPs waveguides (see Fig. ~\ref{fig1}). The input/middle/output SPPs waveguide has the $w = 50 \mu m$, $p = 50 \mu m$, $h = 40 \mu m$ and $a = 20 \mu m$, as shown in Fig. ~\ref{fig1} (b). 
It is already known that the period structure of SPPs waveguide can confine the SPPs in the THz region \cite{Zhang2018, Liu2014} and the SPPs’ transfer on curved waveguide is almost identical to straight case \cite{Pandey2013, Cui2013}, therefore, we can ignore the difference of the coupling strength induced by the curved configuration.
The input/output SPPs waveguide is made of metallic layer and has the special curved configuration. 
The middle SPPs waveguide is the straight configuration with the sandwiched structure (the the ITO layer, the silicon layer and metallic layer), as shown in Fig. ~\ref{fig1} (a). 

The coupling strength between two adjacent SPPs waveguides exponentially decreases along with the distance between two SPPs waveguide and this relationship is consist with coupled mode theory (CMT) \cite{Huang20201, Huang20191, Haus1991} and simulation results (see Fig.~\ref{fig5}). 
The special-curved parameters of the SPPs waveguide coupler are total device length $L = 4000 \mu m$, circular radius $R = 18 mm$, the minimum distance between input/output and middle waveguide $d_{\text{min}} = 60 \mu m$ and distance mismatch $\delta = 1000 \mu m$, as shown in Fig.~\ref{fig1} (c). 
The special curved configuration of input/output SPPs waveguide comes from a well-known coherent quantum control, called Stimulated Raman adiabatic passage (STIRAP) which has already shown the application in the SPPs waveguide coupler \cite{Huang2018, Huang20201, Huang20191}. 

When there is no external voltage on the metal of middle SPPs waveguide ('on-state'), the ITO layer has very high conductivity considering as the gold layer. 
Thus the energy of SPPs can completely transfer from input to output SPPs waveguide via middle SPPs waveguide.
When we apply the external voltage on the metal of middle SPPs waveguide ('off-state'), the ITO layer becomes ultra-low conductive. Therefore, the middle waveguide can not confine the SPPs, in other words, the coupling only happens between top and bottom SPPs waveguides. The distance between top and bottom SPPs waveguides is much larger comparing with 'on-state'. 
Thus, the energy of SPPs hardly transfer from top to bottom SPPs waveguide and confines in the input waveguide along with propagation. 

\begin{figure} [htbp]
\centering
\includegraphics[width=0.5\textwidth]{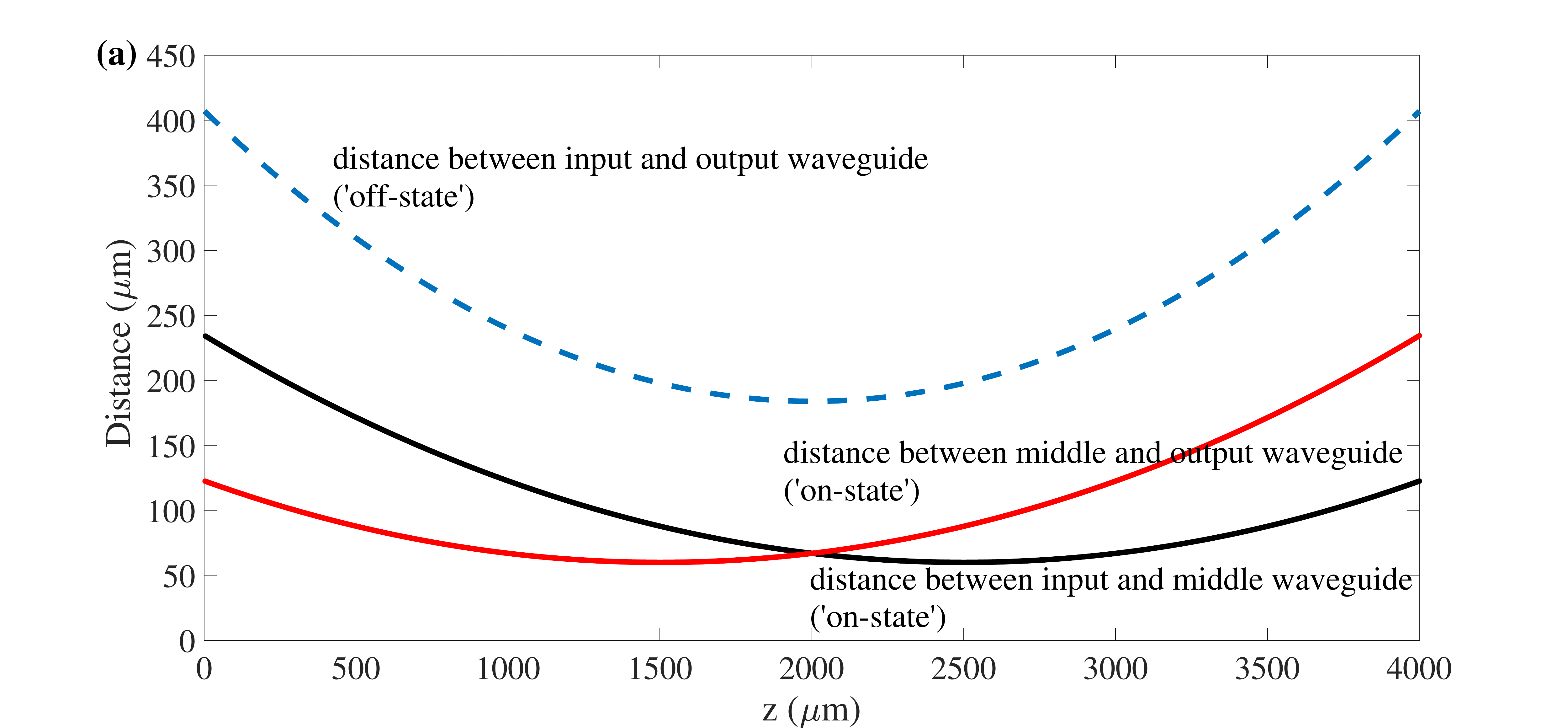}
\includegraphics[width=0.5\textwidth]{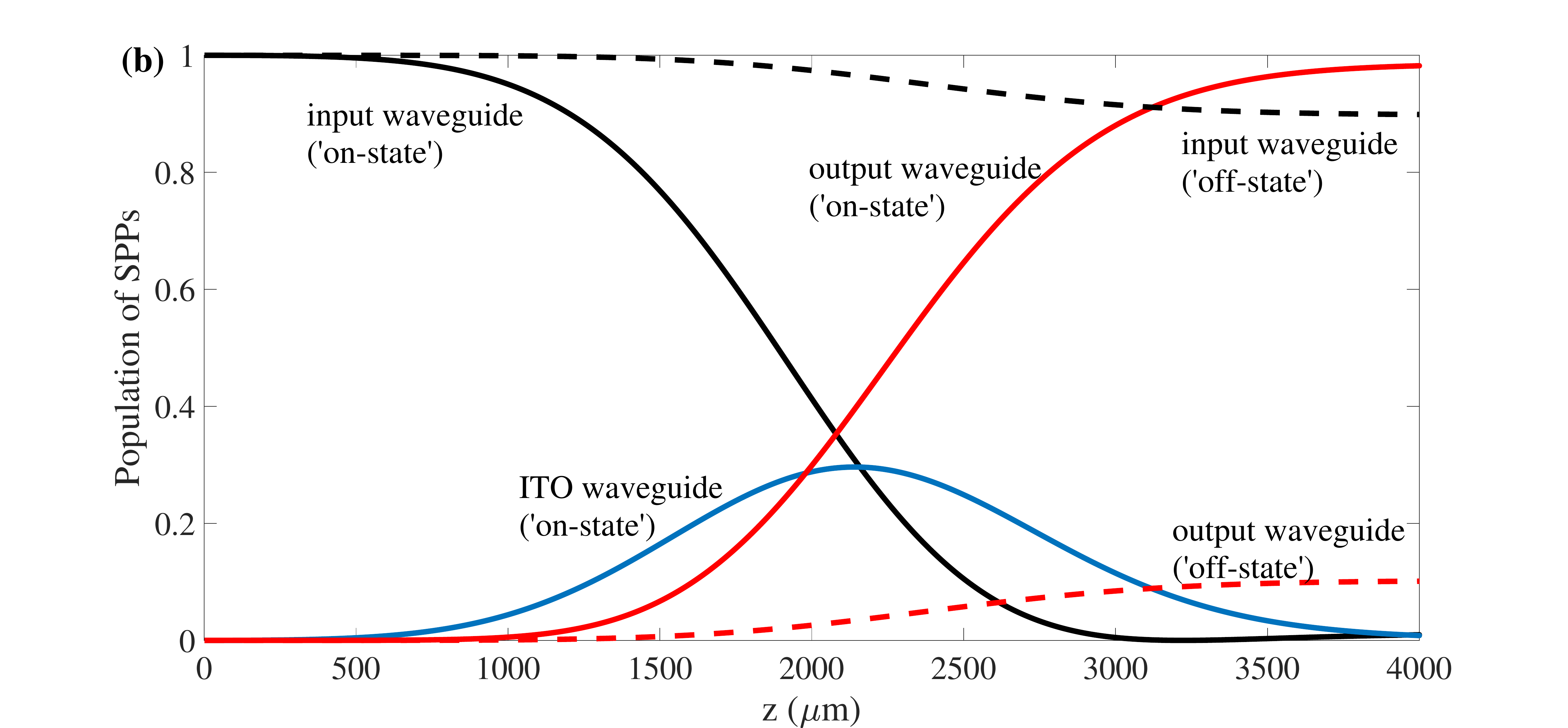}
\caption{(a) The distance configuration of our device with respect to 'on-state' and 'off-state'. The black (red) solid line is the 'on-state' of distance between input and middle (middle and output) waveguide. The dashed blue line is the 'off-state' of distance between input and output waveguide. (b) The corresponding population evolution of SPPs according to 'on-state' and 'off-state'. The solid and dashed lines are with respect to 'on-state' and 'off-state'. The black/red/blue lines are the input/output/middle (ITO) waveguides' energy evolution of SPPs respectively.} 
\label{fig2}
\end{figure} 

In order to demonstrate our idea, we present the distance configuration for 'on-state' and 'off-state' respectively, as shown in Fig.~\ref{fig2} (a).  In the 'on-state', the ITO waveguide with high conductivity can transfer the SPPs and the input and output waveguide can have the coupling via middle ITO waveguide. The black (red) solid line is the 'on-state' of distance between input and middle (middle and output) waveguide. In the 'off-state', the ITO waveguide with very low conductivity can not confine and hold the energy SPPs. Therefore, there is only the direct coupling between input and output waveguide. 
After that, it is already shown that the energy of SPPs evolves along with propagation $z$, which is satisfied with SPPs waveguide coupling equation \cite{Huang20201,Huang20191}, such as
\begin{equation}
i\dfrac{d}{d z}
\begin{bmatrix}
a_{1} \\
a_{2} \\
a_{3}
\end{bmatrix}
= \begin{bmatrix}
0 & C_{1} & 0 \\
C_{1}  & 0 & C_{2} \\
0 & C_{2} & 0
\end{bmatrix} \begin{bmatrix}
a_{1} \\
a_{2} \\
a_{3}
\end{bmatrix},
\end{equation}
where $a_{1}$, $a_{2}$ and $a_{3}$ are the SPPs amplitude of input, middle and output waveguides. The power of input/middle/output SPPs waveguide are $P_{1,2,3} = |a_{1,2,3}|^2$. $C_1$ and $C_2$ are the coupling strengths of input/middle and middle/output SPPs waveguides. Therefore, we can numerically calculate the evolution of SPPs energy along with the propagation based on the coupling strengths between adjacent SPPs waveguides, as shown in Fig.~\ref{fig2} (b). As we can see that the energy of SPPs can completely transfer (more than 98 $\%$) from input to output waveguide via middle waveguide (see the solid lines in Fig.~\ref{fig2} (b)) in the 'on-state' configuration. In the 'off-state' configuration, there is only few energy (less than 8 $\%$) of SPPs transfer from input to  output waveguide, due to the low direct coupling strength between input and output waveguide with low conductivity of middle SPPs waveguide. 
\begin{figure} [htbp]
\centering
\includegraphics[width=0.5\textwidth]{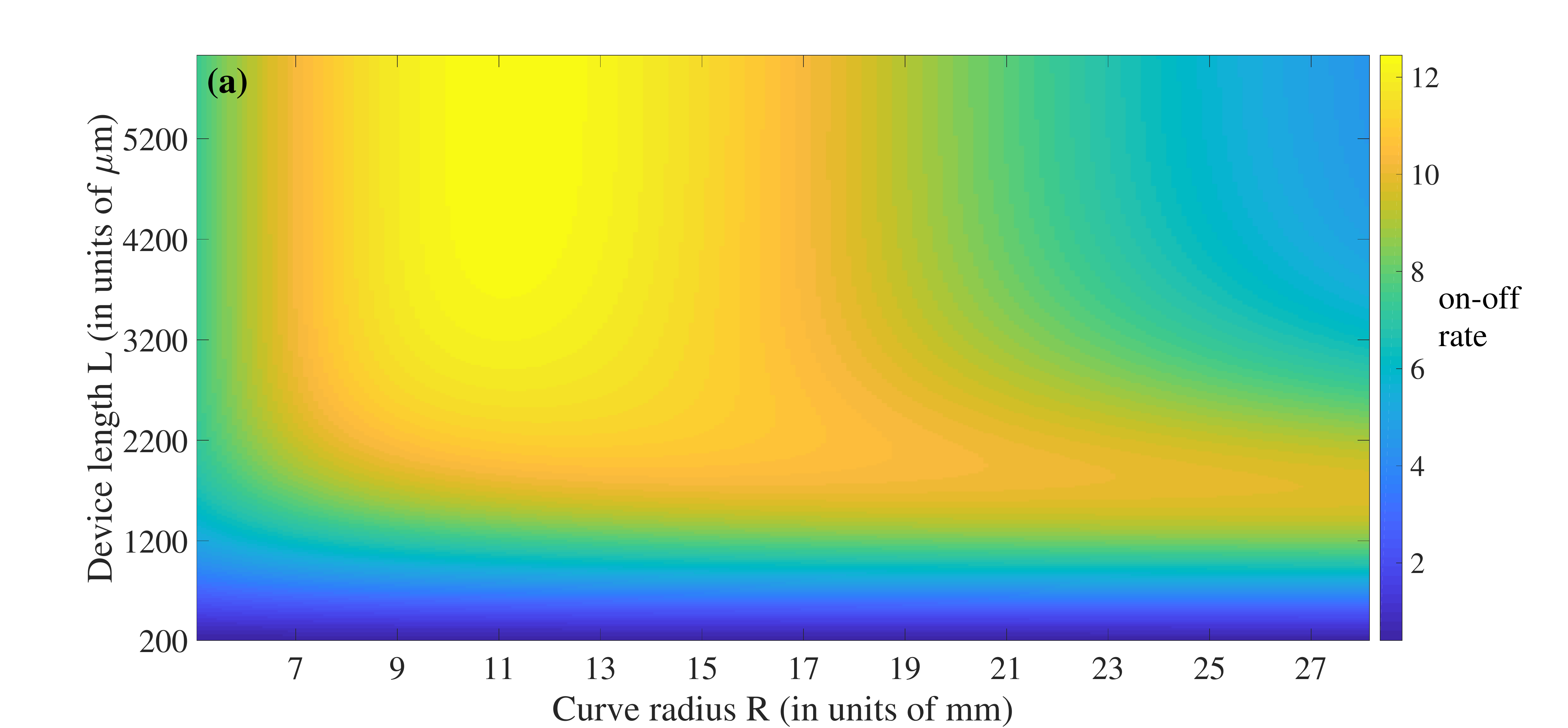}
\includegraphics[width=0.5\textwidth]{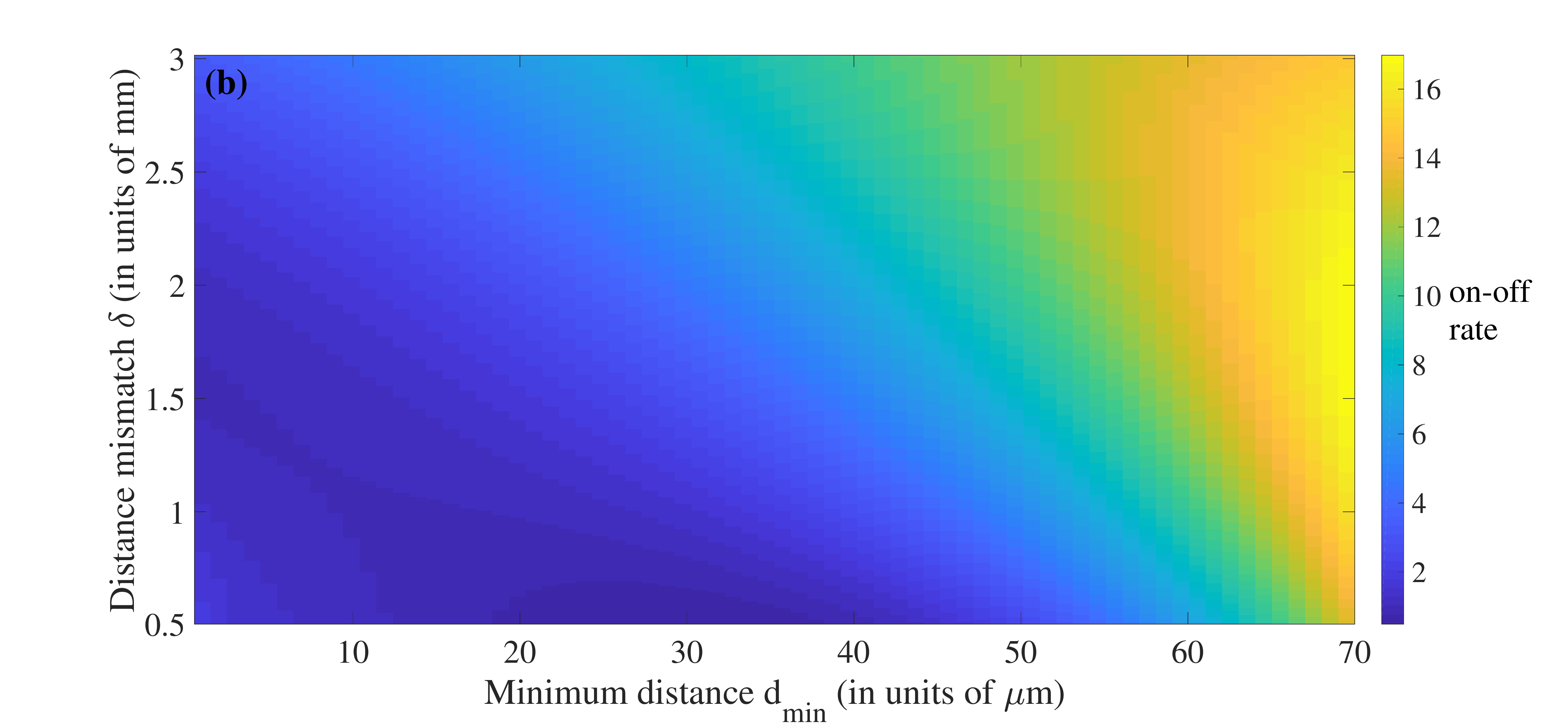}
\caption{The on-off rate varying with different geometrical parameters. (a) The contour plot of on-off rate with varying the device length $L$ and curve radius $R$. (b) The contour plot of on-off rate with varying the distance mismatch $\delta$ and minimum distance $d_{\text{min}}$.} 
\label{fig3}
\end{figure} 

\section{Robust and broadband switch}
%***************************************************************
In this section, we demonstrate that our device is the robustness against the geometrical parameters length $L$,  curve radius $R$, distance mismatch $\delta$ and minimum distance $d_{\text{min}}$ by using the SPPs coupler equations for both 'on-state' and 'off-state'. Robustness against the geometrical parameters is a large advantage for the device fabrication, which largely increases the error suffering during the processing of fabrication and robustness hugely decreases the cost of device, due to not requirement of high precise equipment. In order to measure the performance of our switch in the same configuration, we introduce the on-off rate with $P_{\text{on}}/ P_{\text{off}}$, which the 'on-state' divided by 'off-state' of final SPPs energy of output waveguide to measure the energy difference of 'on-state' and 'off-state'. Our 'on-state' and 'off-state' are separated by input and output waveguides (two different light pathway), therefore, if on-off rate is higher than 10, we can easily separate 'on-state' and 'off-state' by SPPs energy difference. 
The Fig.~\ref{fig3} demonstrates the on-off rate against varying different geometrical parameters. As we can see from the Fig.~\ref{fig3}, we can easily obtain a large and continuous area of high on-off rate (higher than 10) with different geometrical parameters. Therefore, we can claim that our device is robust against the perturbations of geometrical parameters (such as the device length $L$, curve radius $R$, distance mismatch $\delta$ and minimum distance $d_{\text{min}}$).

\begin{figure*} [htbp]
\centering
\includegraphics[width=0.6\textwidth]{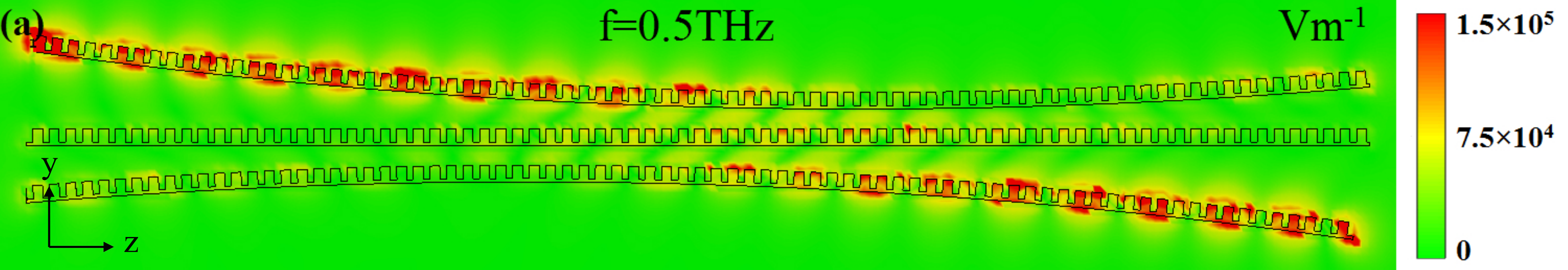}
\includegraphics[width=0.6\textwidth]{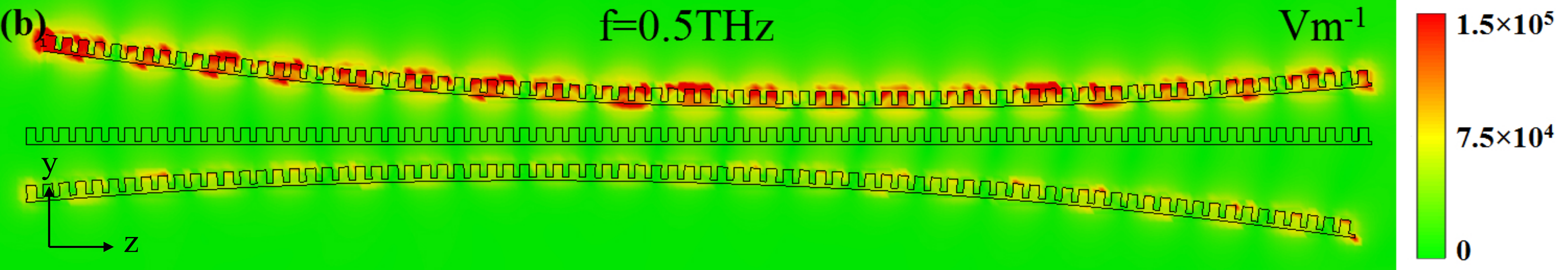}
\includegraphics[width=0.6\textwidth]{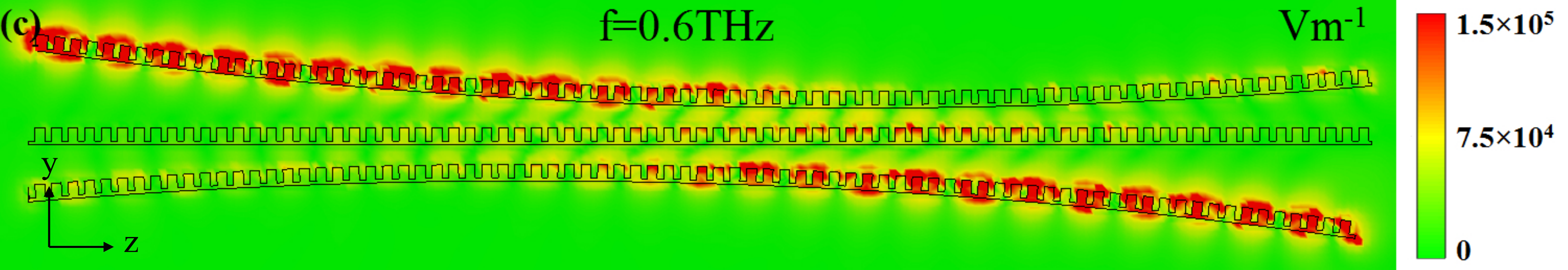}
\includegraphics[width=0.6\textwidth]{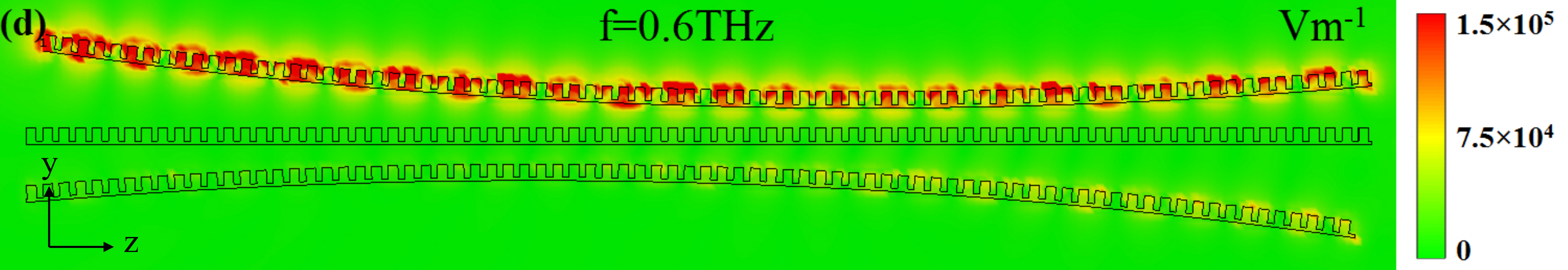}
\includegraphics[width=0.6\textwidth]{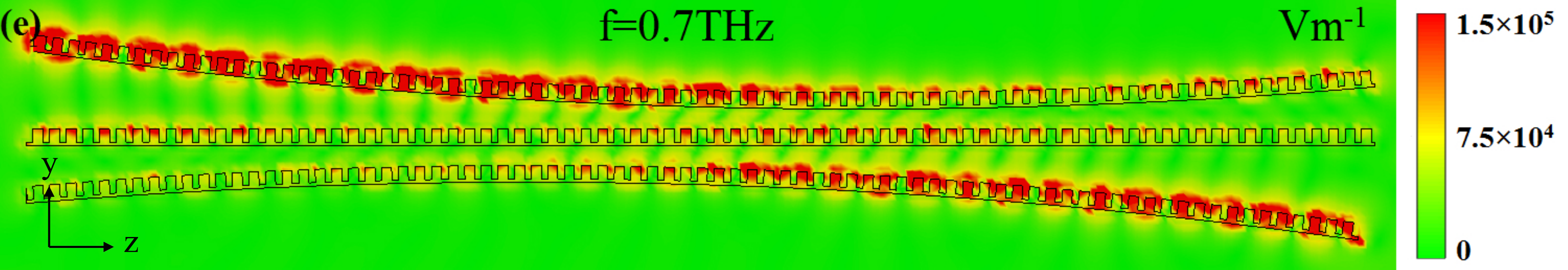}
\includegraphics[width=0.6\textwidth]{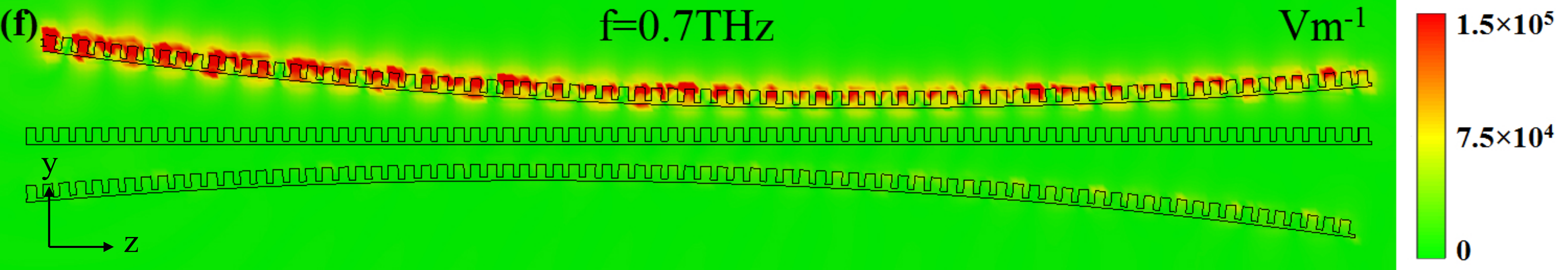}
\caption{The 'on-state' (a, c, e) and 'off-state' (b, d, f) of our switch with (a, b) f = 0.5 THz, (c, d) f = 0.6 THz and (e, f) f = 0.7 THz. } 
\label{fig4}
\end{figure*} 

Furthermore, we employ the numerical software to full wave simulate the our switch based on different input frequencies. Fig.~\ref{fig4} (a) (c) (e) are the 'on-state' for different frequencies and remaining figures are the 'off-state' corresponding to the frequencies. As we can see that our device have a good performance both on 'on-state' and 'off-state' at frequencies from 0.5 THz to 0.7 THz in our this specific example configuration.
When the frequencies are smaller than 0.5 THz, the 'off-state' of our device do not have a good performance and too much power can be transferred to output waveguide, because the coupling strength becomes larger along with frequency decreasing with large gap (see the appendix Fig.~\ref{fig5}). As we can see from the appendix Fig.~\ref{fig5}, the coupling strength becomes much smaller at 0.8 THz when the gap is 60 $\mu m$. Therefore, the 'on-state' of our device are not good at frequencies larger than 0.7 THz and coupling strengths are too small to support SPPs energy from input to middle SPPs waveguide. 

\begin{figure} [htbp]
\centering
\includegraphics[width=0.5\textwidth]{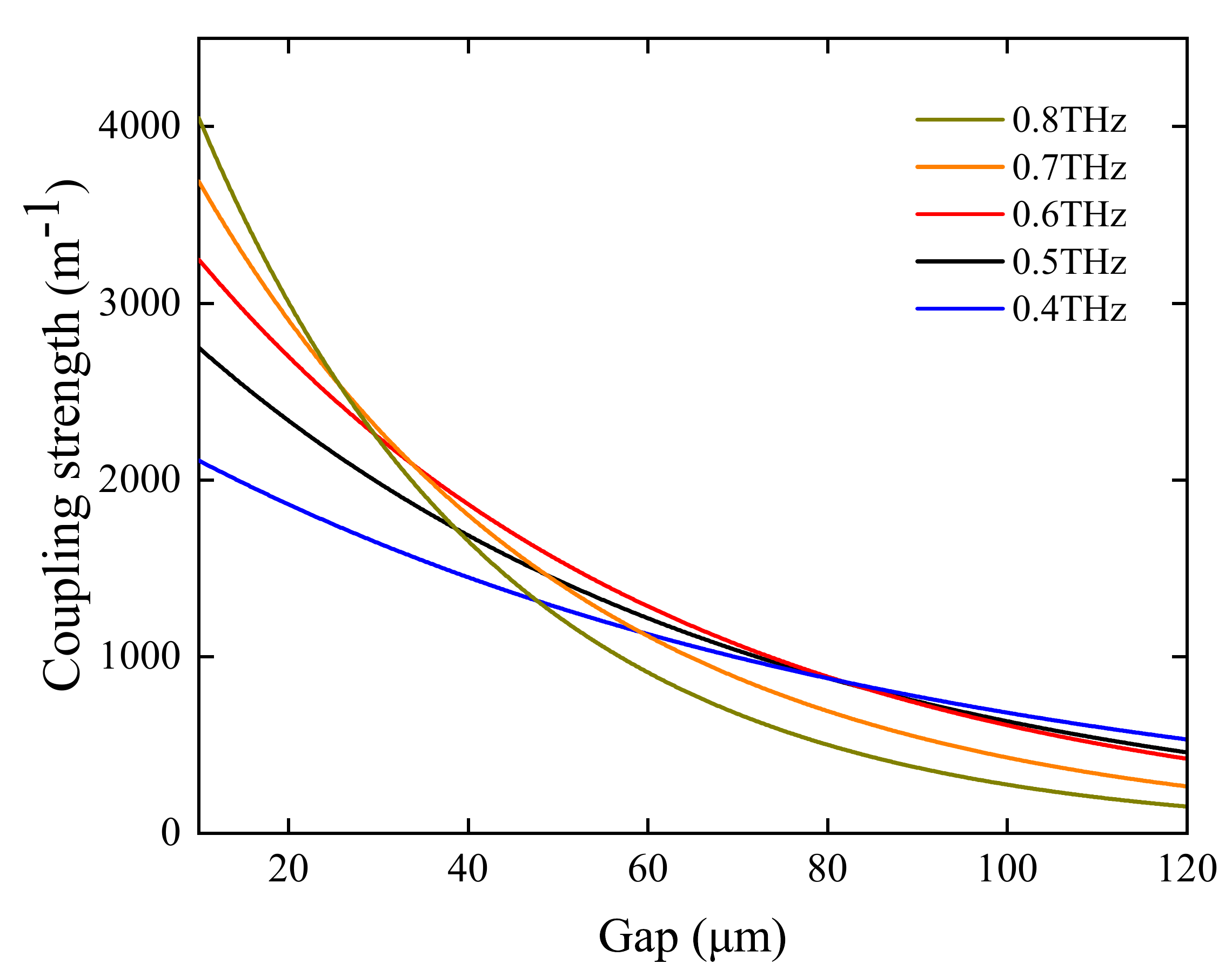}
\caption{The relationship of coupling strength and the gap between two SPPs waveguides.} 
\label{fig5}
\end{figure} 

\section{Conclusion}
In this paper, we propose a novel design of broadband and robust THz SPPs switch based on waveguide coupler embedded the coherent quantum control. 
From our example configuration, our switch is robust against the geometrical parameters, including the device length $L$, curve radius $R$, distance mismatch $\delta$ and minimum distance $d_{\text{min}}$. Furthermore, we demonstrate the broadband performance of our switch from 0.5 THz to 0.7 THz for both 'on-state' and 'off-state'. We believe this funding will be the great improvement for the integrated optical computation and THz communication. 

\section*{Appendix}

Due to the complicated boundary conditions, it is very hard to obtain the mode profile of the SPPs waveguides in the analytical solution, thus, analytical solving the relationship between coupling strength and gap between two SPPs waveguides is complicated. Therefore, we employs the full-wave simulation of two parallel SPPs waveguides to obtain the relationship between coupling strength and gap between two SPPs waveguides (see Fig.~\ref{fig5}), which is consistent with the exponential relationship by coupled mode theory predicting \cite{Haus1991}.

\section*{Acknowledgements}
This work acknowledges funding from National Key Research and Development Program of China (2019YFB2203901); National Science and Technology Major Project (grant no: 2017ZX02101007-003); National Natural Science Foundation of China (grant no: 61565004; 61965005; 61975038; 62005059; 62075248); National Key R$\&$D Program of China (Grant No. 2017YFB0405400; 2020YFB2001903). the Science and Technology Program of Guangxi Province (grant no: 2018AD19058). W.H. acknowledges funding from Guangxi oversea 100 talent project; W.Z. acknowledges funding from Guangxi distinguished expert project.

%***************************************************************

\end{document}